\documentclass[twocolumn,showpacs,preprintnumbers]{revtex4}
\usepackage{amssymb}
\usepackage{graphicx}

\begin{document}

\title{Effects of dipole-dipole interaction between cigar-shaped BECs of
cold alkali atoms: Towards inverse-squared interactions}
\author{Yue Yu}
\affiliation{Department of Physics and Center for Field Theory and Particle Physics,
Fudan University, Shanghai 200433, China}
\affiliation{The State Key Laboratory of Theoretical Physics, Institute of Theoretical
Physics, Chinese Academy of Sciences, P.O. Box 2735, Beijing 100190, China}
\author{Zhuxi Luo}
\affiliation{School of Economics, Fudan University, Shanghai 200433, China}
\author{Ziqiang Wang}
\affiliation{Department of Physics, Boston College, Chestnut Hill, MA 02467, USA}
\date{\today }

\begin{abstract}
We show that the dipole-dipole coupling between Wannier modes in
cigar-shaped Bose-Einstein condensates (BECs) is significantly enhanced
while the short-range coupling strongly suppressed. As a result, the
dipole-dipole interaction can become the dominant interaction between
ultracold alkali Bose atoms. In the long length limit of a cigar-shaped BEC,
the resulting effective one-dimensional models possess an effective inverse
squared interacting potential, the Calogero-Sutherland potential, which
plays a fundamental role in many fields of contemporary physics; but its
direct experimental realization has been a challenge for a long time. We
propose to realize the Calogero-Sutherland model in ultracold alkali Bose
atoms and study the effects of the dipole-dipole interaction.
\end{abstract}

\pacs{67.85.Pq, 37.10.Jk, 11.30.Pb}
\maketitle

\section{Introduction}

The quantum simulation of strongly correlated systems with cold atoms
and molecules is an advancing subject in modern physics \cite{ho}. With
precise control of system parameters by external fields, the quantum simulation becomes an
ideal method to capture the key physics of such formidable many-body
systems. The first example is the simulation, using boson atoms, of the
Bose-Hubbard model \cite{ho}, the most basic and fundamental model of
strongly correlated electron materials that remains to be better understood.
Other examples of quantum simulation include using fermion atoms to simulate the fermionic
Hubbard model \cite{hubbard}, polar molecules and Rydberg atoms to simulate
various quantum spin and more exotic quantum systems \cite{pm,rd}.

Among the strongly correlated models of condensed matter and many-body
systems, there is an important class of one-dimensional models with
inverse-squared interactions. The prototype of the latter is the celebrated
Calogero-Sutherland model \cite{ca,suth} that nowadays plays a
fundamental role in many branches of modern physics \cite{csmb,pol}. Besides
its importance in theoretical and mathematical physics, the Calorego-Sutherland type of
models are often applied to explain the physical phenomena in fractional
quantum Hall effects \cite{yu}, quantum exclusion statistics \cite{berwu},
black hole physics \cite{bh}, quantum spin systems \cite{xxx}, and so on. It
is of great interests to directly simulate the Calorego-Sutherland model using ultracold
atoms. However, the inverse-squared interaction presents a serious challenge
for experimental realizations amid the lack of a concrete proposal.

In this paper, we propose a possible realization of the Calorego-Sutherland type of models
by suppressing the $s$-wave interaction while enhancing the dipole-dipole
 interaction of cold alkali atoms on optical lattices. The dipole-dipole
interaction between cold particles \cite{yl} is the basis for the cold polar
molecular simulator and Rydberg simulator. These cold particle gases usually
have strong dipolar couplings \cite{pmole,dpdp,dpdp1,dpdp2,rd}. For alkali
atoms, the effect of the dipole-dipole interaction was thought to be of negligible
significance because the dipole-dipole coupling is too weak compared to the $s$-wave
scattering. However, the cold alkali atom gases are much easier to be made
and manipulated, and are far more stable than the polar molecule and dipolar
atomic gases.

We study the effective interactions between the
single mode fluctuations in an array of cigar-shaped cold alkali atom
Bose-Einstein condensates (BECs) which are confined in a one-dimensional
optical lattice \cite{stoof,van}. Remarkably, we find that the dipole-dipole interaction in these systems reduces to an
inverse-squared interaction between the single modes with a significantly
enhanced coupling strength proportional to the number of atoms in a single
BEC. Moreover, the $s$-wave scattering is strongly reduced with the
effective short-range interaction strength scaling approximately as the
inverse of the number of atoms. As a result, in the limit where the single
BEC is long, the effective on-site interaction can be neglected and the
system is governed by the dominating inverse-squared interaction, i.e., the
Calorego-Sutherland model. The material parameters of the alkali metals will be considered
for the feasibility of our proposal of a quantum simulator of the Calorego-Sutherland model
to study the dipole-dipole interactions.

This paper was organized as follows: In Sec. II, we setup the system we are considering and  derive the effective single-band model.  In Sec. III, we show that why the inverse-square potential becomes dominant in this system and  study the dilute gas limit.
The section IV demonstrate the experimental implications. We conclude in Section V.

\begin{figure}[tbp]
\begin{center}
\includegraphics[width=5cm]{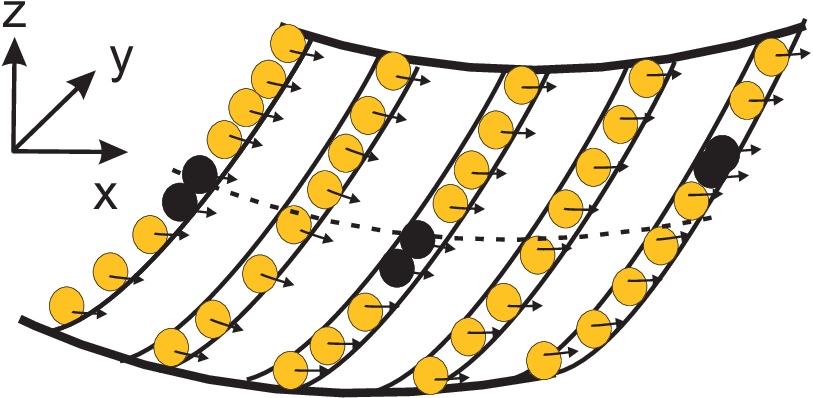}{\small (a)}
\end{center}
\vspace{3mm}
\begin{center}
\includegraphics[width=5cm]{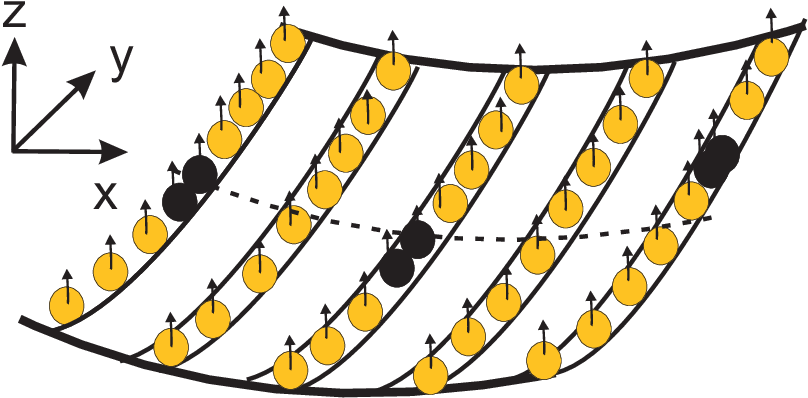}{\small (b)}
\end{center}
\caption{(color online) The array of cigar-shaped cold atom clouds on an
optical lattice. (a) The dipoles are polarized along the $x$-direction. (b)
The dipoles are polarized along the $z$-direction. The yellow spots are the
BEC substrate and the black spots are the quasi-atom with the
inverse-squared interaction. }
\label{fig1}
\end{figure}
\begin{figure}[tbp]
\begin{center}
\includegraphics[width=9.0cm]{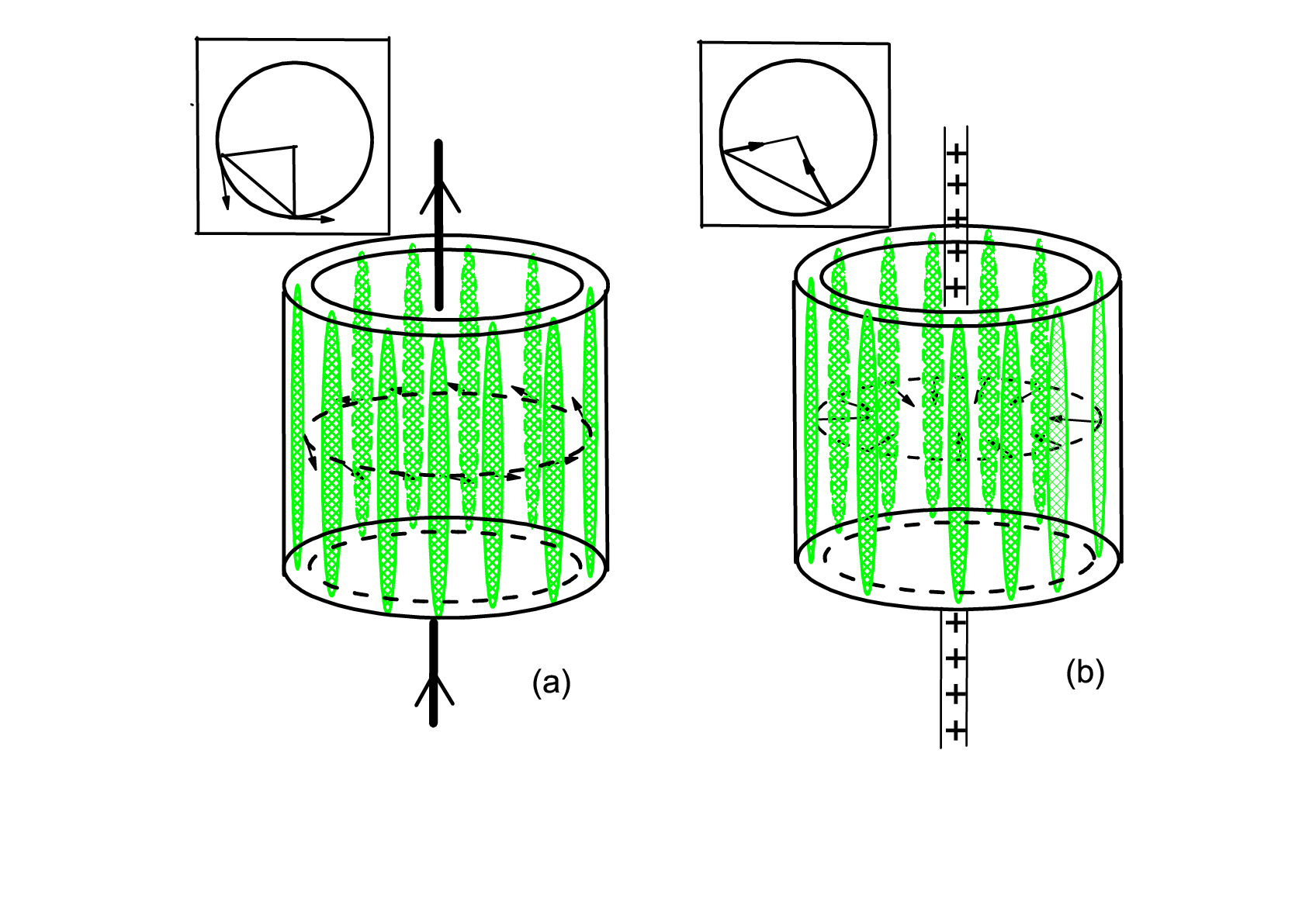}
\end{center}
\par
\vspace{-0.5cm}
\caption{(color online) Cigar-shaped BEC (green) confined to a cylinder. (a)
The magnetic dipoles; (b) The electric dipoles. The inserts indicate the
directions of the dipoles.}
\label{fig2}
\end{figure}

\section{Effective single-band model}

\subsection{Setups of the system}

We consider a cold Bose alkali atom cloud with a small dipole moment ($\sim
1\mu_B$). In the presence of a strong $z$-direction confinement, the cloud
has a pancake shape trapped in the $x$-$y$ plane with the trapping
frequencies $\omega_x$ and $\omega_y$. The size of the system is therefore $L_{x,y}=\sqrt{\hbar/m\omega _{x,y}}$. Using a periodic optical potential
along the $x$-direction, the pancake is incised into a lattice of
cigar-shaped gases. The frequency of the optical potential is denoted as $\omega_p$, with the oscillator length satisfying $\hbar \omega _{p}=\hbar$. An applied external field parallel to the $x$-axes, or
perpendicular to the $x$-$y$ plane polarizes all the dipoles along the field
direction as shown in Fig.\ref{fig1}(a) and (b), respectively. This setup is
towards the Calogero model \cite{ca}.

Another setup follows from rolling the $x$-$y$ plane to form the surface of
a cylinder as shown in Fig.\ref{fig2}. This geometry of the optical lattice
is not yet realized. One possibility is using the Laguerrer-Guassian laser
beams, which have been applied or proposed to build many periodic optical
lattices, e.g., the circular lattice \cite{cir}, the ring lattice \cite{ring}%
, and even a possible torical lattice \cite{tor}. However, there are many technical difficulties and
complication need to be engineered.  It will not be the main focus of this work.
Sutherland's periodic variant \cite{suth} can be realized with this setup.

\subsection{Derivation of an effective single-band model}

We first focus on the setup shown in Fig. \ref{fig1} and set $\omega_x=0$.
The energy scale in the problem is
\[
\hbar\omega_y\ll k_BT\ll \hbar\omega_p.
\]
Thus, we are faced with a multi-band problem. Following the
two-step procedure proposed in \cite{van}, we aim to reduce the multi-band
problem to an effective single-band one. First, we treat a single
cigar-shaped gas on a given lattice site. The lowest energy level is a BEC and higher levels correspond to the thermal cloud. Denote on every site the wave function of a whole cigar-shaped BEC $%
\Psi_0(y)$, and the number of BEC atoms on each site $N_0$. In addition, the condensate size is marked as $R_y$ on each site. The equation
of state with only short-range interactions has been solved exactly in the
weak interaction limit \cite{stoof}. It was shown that even in the presence
of phase fluctuations (collective modes), the atom number in a BEC can be
determined at any given temperature \cite{van}. For atoms carrying a dipole,
in addition to the short-range interaction, there is a dipole-dipole interaction
potential
\begin{eqnarray}
V_d(\mathbf{r},\mathbf{r}^{\prime})=\frac{\mathbf{d}\cdot \mathbf{d}%
^{\prime}-3(\mathbf{d}\cdot\hat{\mathbf{R}})(\mathbf{d}^{\prime}\cdot \hat{%
\mathbf{R}})}{R^3},
\end{eqnarray}
where $\mathbf{R}=\mathbf{r}-\mathbf{r}^{\prime}$ and $\hat{\mathbf{R}}=%
\mathbf{R}/R$; $\mathbf{d}$ and $\mathbf{d}^{\prime}$ are the dipole moments
of the atoms located at $\mathbf{r}$ and $\mathbf{r}^{\prime}$,
respectively. We consider all $\mathbf{d}$ to be parallel as shown in Fig. %
\ref{fig1}. The Fourier components of the dipolar interacting potential do
not contain any singularity in momentum space other than providing an
anisotropic term. Therefore, for a cold atom gas with weak dipolar
interactions, this anisotropy does not cause a qualitative difference and we
can continue to use the method given in Ref. \cite{stoof} to solve the
equation of state and determine the atom number $N_0(T)$ in a BEC.

The second step is to consider the coupling between sites \cite{van}. If $%
\hbar\omega_y\ll N_yU_0 \ll \hbar\omega_p$, where $N_y$ is the average atom
number in a cigar-shaped gas(including those in thermal cloud) and $U_0$ the on-site repulsion, the wave
function may be approximated by the product of the single atom ground state
wave function in the $x$-direction and the cigar-shaped atom gas in the $y$
direction. Notice that because we have a BEC $\Psi_0(y)$ at every site and obviously $%
N_0\lesssim N_y$, the coupling between the sites will be dominated by
tunneling from a BEC to a condensate as opposed to a thermal cloud. As a
result, this multi-band problem can be reduced to a single band one \cite%
{van}, i.e., the field operator can be approximated by
\[
\psi(\mathbf{r})\approx \sum_{i} a_{i} w(x-x_i)\Psi_0(y),
\]
where the bosonic mode $a_i$ corresponds to annihilating a quasi-atom mode
described by the product of the Wannier function $w(x)$ and the cigar-shaped
BEC wave function $\Psi_0(y)$. Under this approximation, we arrive at an
effective single band boson model with renormalized on-site and renormalized
dipole-dipole interactions \cite{van}.

\section{The lattice model with inverse-squared interaction}

\subsection{Inverse-squared Interaction}

Keeping only the nearest neighbor hopping and expanding the on-site energy
to the second order near the average occupation $N_{0}$ \cite{van,li}, we
obtain the effective single-band lattice model described by the Hamiltonian,
\begin{eqnarray}
H &=&-\sum_{\langle ij\rangle }ta_{i}^{\dag }a_{j}+\frac{U_{R}}{2}%
\sum_{i}\delta n_{i}(\delta n_{i}-1)  \nonumber \\
&+&\sum_{i<j}U_{d,ij}\delta n_{i}\delta n_{j},  \label{lh}
\end{eqnarray}%
where $\delta n_{i}=a_{i}^{\dag }a_{i}-N_{0}$ is the deviation of the atom
number from the average number per site; $t$ is the nearest neighbor hopping
amplitude. We emphasize that here the hopping 'particle' is not the whole
cigar-shaped BEC as a 'giant particle' but only the particle number variance
between the adjacent lattice sites. We define the original bare on-site repulsion as
\[
U_{0}=\frac{4\pi \hbar ^{2}a}{m}\int dxdydz|\psi _{0}(z)|^{4}|w(x)|^{4}|\psi
_{0}(y)|^{4}
\]%
where and $a$ is the $s$-wave scattering length, $\psi _{0}(z)$ is the
perpendicular confined single particle wave function and $\psi _{0}(y)$ is
the single particle wave function in the $y$-direction.

The renormalized on-site potential $U_{R}$ due to the condensation of atoms in the $y$-direction is {\it not simply} given by replacing
$\psi_0(y)$ in $U_0$ by $\Psi_0(y)$, the condensate wave function.  Because the condensed wave function is  spread out by the on-site repulsion, $U_R$   is defined by \cite{van}
\begin{eqnarray}
U_R=\frac{\partial^2 F}{\partial N_0^2},
\end{eqnarray}
where $F$ is on-site free energy given by \cite{li}
\begin{eqnarray}
F=F_{TF}+\int dx w^*(x)(-\hbar^2\frac{d^2}{dx^2}+V(x))w(x).
\end{eqnarray}The Thomas-Fermi free energy $F_{TH}$ is the Gross-Pitaevskii energy of the $y$-direction one-dimensional BEC
in which the dipole-dipole interaction within the single BEC is considered.

The ratio between $U_R$ and $U_0$ without dipole-dipole
interaction has been estimated in Ref. \cite{van}, $U_R=U_{0}\frac{l_{p}}{R_{y}}$,
where $R_{y}$ is the Thomas-Fermi radius of the cigar-shaped BEC in the $y$%
-direction and $l_{p}$ is the lattice spacing under the consistency
condition
\[
L_{y}/a\ll N_{0}\ll (\hbar \omega _{p}/\hbar \omega _{y})^{2}\sqrt{2\pi }%
l_{p}/a.
\]%
 Hereafter, we set $l_{p}=1$ as the unit of length unless
stated explicitly.  The bare on-site interaction is strongly renormalized because the on-site wave function spreads out the BEC wave function \cite{van}.

For alkali atoms we are studied, the bare dipole-dipole interaction in a single cigar-like BEC  is very weak comparing to $U_0$.
Therefore, we can take $U_R\sim U_0\frac{l_p}{R_y}$ as a good approximation. The only change is the Thomas -Fermi radius $R_y$ which is determinted when the dipole-dipole interaction between the atoms within the single BEC is included.

The hopping amplitude $t$, as pointed out by Refs. \cite{van,li}, is almost not renormalized because  the transverse Wannier function  $w(x)$ is almost not affected by the spread-out of the BEC wave function.

The last term in Eq.~(\ref{lh}) is the renormalized
dipolar interaction potential between the fluctuating modes of different cigar-like condensates ($i<j$). This is  given by
\begin{eqnarray}
U_{d,ij} &=&\int d\mathbf{r}d\mathbf{r}^{\prime
}|w(x-x_{i})|^{2}|w(x^{\prime }-x_{j})|^{2}  \nonumber \\
&&V_{d}(\mathbf{r,r}^{\prime })|\Psi _{0}(y)|^{2}|\Psi _{0}(y^{\prime
})|^{2}.
\end{eqnarray}%
Consider the case where the condensate size $R_{y}$ is much larger than the
lattice size $L_{x}$, i.e. $R_{y}\gg L_{x}$. We approximate the BEC wave
function by the average density of atoms in a single BEC, $|\Psi
_{0}(y)|^{2}\approx N_{0}/R_{y}$. Under this condition, it is easy to carry
out the integration over $y$ and $y^{\prime }$ and arrive at
\begin{eqnarray}
&&U_{d,ij}\approx \int dxdx^{\prime }|w(x-x_{i})|^{2}|w(x^{\prime
}-x_{j})|^{2}  \label{2} \\
&&\times \frac{GR_{y}}{(x-x^{\prime })^{2}\sqrt{(x-x^{\prime })^{2}+R_{y}^{2}%
}}\approx \frac{G}{(x_{i}-x_{j})^{2}}+O(\frac{1}{R_{y}^{2}}),  \nonumber
\end{eqnarray}%
where $G=-2d^{2}N_{0}^{2}/R_{y}$ for the setup shown Fig. \ref{fig1}(a) and $%
G=2d^{2}N_{0}^{2}/R_{y}$ for the setup in Fig. \ref{fig1}(b). Thus, we have
obtained a one-dimensional lattice model with on-site and inverse-squared
interactions between the single modes in the ground state. The finite size
effect of the cigar-shaped BEC only contributes a negligible $O({1}/{%
R_{y}^{2}})$ correction. We observe that while the on-site coupling constant
in one dimension is reduced a factor $l_{p}/R_{y}$, the long range
interacting coupling is renormalized by an additional factor $2N_{0}^{2}$
besides the reduced factor $l_{p}/R_{y}$. If $N_{0}^{2}\gg R_{y}/l_{p}$, the
latter can be remarkably enhanced.

\subsection{Dominance of the renormalized dipole-dipole interaction}

The stability condition of a single BEC at a given site is governed by the
ratio of dipolar to on-site interactions $\epsilon _{dd}=d^{2}/(3U_{0})<1$
\cite{dpt2}, where $d$ is the magnitude of the dipole moment. For alkali
metal atoms, $d\sim 1\mu _{B}$ and $\epsilon _{dd}\ll 1$. For example, $%
\epsilon _{dd}=0.007$ for $^{87}$Rb. In the present context, while the
renormalized on-site interaction $U_{R}$ is reduced by a factor $\frac{l_{p}%
}{R_{y}}$ from $U_{0}$ \cite{van}, our results show that the effective
dipole moment is enhanced to $d_{R}=dN_{0}$ as shown in Eq.~(\ref{2}).
Therefore, although the bare ratio $\epsilon _{dd}\ll 1$, the renormalized
on-site interaction can be negligibly weak compared to the renormalized
dipole-diploe interaction in the effective one-dimensional single-band
model. Indeed, it is instructive to introduce a renormalized ratio of
dipolar and on-site interactions, $\epsilon
_{d_{R}d_{R}}=d_{R}^{2}/(3U_{R})=\epsilon _{dd}N_{0}^{2}R_{y}/l_{p}$. The
latter that can be much larger than unity. Assuming $N_{0}\sim 10^{3}$ and $%
R_{y}/l_{p}\sim 10^{3}$, we estimate that $\epsilon _{d_{R}d_{R}}\gtrsim
10^{6}$ for $^{87}$Rb.

\subsection{Trapping potential in the $x$-direction}

We have obtained a homogeneous lattice boson model with inverse-squared
interaction by neglecting the trapping in the $x$-direction. Putting back
the harmonic trap with the frequency $\omega_x$ may affect the atom number $%
N_0(x)$ at each site. If the width of a single BEC is too wide, the number
of atoms per site may vary from site to site and $N_0$ in the center of
trapping potential can be several times that far away from the center.
However, within a cigar-shaped dipolar BEC, the dipole-dipole repulsion is strong
enough against increasing the width of the single BEC at the trapping
center. Therefore, the factorization of the ground state wave function is
still valid and $N_0$ remains approximately a constant. The resulting
inhomogeneous model is given by $H$ in (\ref{lh}) augmented by an additional
term $\sum_iV_{h,i} \delta n_i$, where
\begin{eqnarray}
V_{h,i}=\frac{1}2m\omega^2_x \int dx |w(x-x_i)|^2x^2  \label{har}
\end{eqnarray}
is the harmonic trapping potential along the $x$-direction.

\subsection{Dilute gas limit and the Calorego-Sutherland model}

Denoting the total fluctuations in the atom number as $\delta N=\sum_i
\delta n_i$. The fixed number of $\delta N$-particle system may be described
by a continuum model in the dilute gas limit with $\delta N/L_x\ll1$, where
the dispersion $-t\cos k/\hbar\sim k^2/(2m_R)$ with the effective mass $%
m_R\sim \hbar^2/t$. Dropping the negligible on-site interaction as argued
above, the effective Hamiltonian reads
\begin{eqnarray}
H_{CS}=\sum_{i=1}^{ \delta N}\biggl(-\frac{\hbar^2}{2m_R}\frac{d^2}{dx^2_i} +%
\frac{m_R\omega_R^2}2x_i^2\biggr)+\sum_{i< j}\frac{G}{|x_{ij}|^2},  \label{4}
\end{eqnarray}
where $x_{ij}=x_j-x_k$ and the effective trapping potential is defined via $%
m\omega_x^2=m_R\omega_R^2$. The model described by the Hamiltonian (\ref{4})
is precisely the celebrated Calorego-Sutherland model \cite{ca,suth}. A bosonic ``particle''
at $x_i$ in this continuum model corresponds to one more atom than $N_0$ at
site $i$ in the lattice model. The ground state wave function is given by
\begin{eqnarray}
\Psi_{0,\lambda}(x_1,...,x_{\delta N})=\prod_{ 1\leq j<k\leq {\delta N} }
|x_{jk}|^\lambda e^{-\frac{m_R\omega_R}{2\hbar} \sum_jx^2_j }  \label{ground}
\end{eqnarray}
with the ground state energy $E_g=\frac{1}2{\delta N}\hbar\omega_R(\lambda ({%
\delta N}-1)+1)$. Here, $\lambda=\lambda_\pm=\frac{1\pm\sqrt{1+4g}}2$ are
the solutions of the equation $\lambda(\lambda-1)=g=(m_R/\hbar^2)G$. For a
weak on-site interaction with $U_0/t<1$, $g$ may vary as $t$. If $%
l_p/R_y\sim O(10^{-3})$, $N_0\sim O(10^3)$, $U_0/t\sim O(10^{-1})$, the
absolute value of $g$ is of the order unity. For example, for the repulsive
interacting $^{87}$Rb atoms, we estimate $g\approx 0.021N_0^2U_0/tR_y\sim2.1$
\cite{note}. The exact solution of the Calorego-Sutherland model can now be utilized to make
predictions for the stability of the ultracold alkali atoms.

For the setups in Fig. 1(a), $g<0$. If $g<-1/4$, $\lambda$ becomes
imaginary. This leads to an imaginary ground state energy and the system is
unstable. However, when $-1/4\leq g<0$, the ground
state is stable for $0\leq\lambda_-<1/2$ despite of the attractive
interaction.

For the setups in Fig. 1(b), $g>0$. For $-\frac{1}2<\lambda_-<0$, i.e, $g<%
\frac{3}4$, the wave function is still square integrable. This implies that
the system will also unstable if the repulsion between particles is not
strong enough. On the other hand, for $g>\frac{3}4$, (\ref{ground}) with $%
\lambda=\lambda_-<-\frac{1}2$ is not a physical solution because it is not
square integrable. A physically stable solution is given by (\ref{ground})
with $\lambda=\lambda_+$.

We next consider the geometry in Fig. \ref{fig2}. In this geometry, the
cylinder extends along the $y$-direction and circulates in the $x$%
-direction. For $R_{y}\gg L_{x}$, one may prove that, up to a constant term,
the renormalized dipole-dipole interaction is a periodic inverse-squared interaction
as in the Sutherland model \cite{suth}
\[
V\approx \pm \sum_{i\neq j}\frac{\pi ^{2}\lambda (\lambda -1)}{L_{x}^{2}\sin
^{2}[\pi x_{ij}/L_{x}]}+O(\frac{1}{R_{y}^{2}}),
\]%
where $\pm $ are corresponding to Fig. \ref{fig2}(a) and Fig. \ref{fig2}(b),
respectively. The similar but periodic solutions were obtained by Sutherland
\cite{suth}.

\section{ Experimental implications}

The first prediction on the effects of the dipole-dipole interaction between the
alkali atoms is unstable when $g<-1/4$.  The Calogero-Sutrherland gas collapses
the single modes to a single lattice site (a
given single cigar-shaped BEC) when $g<-1/4$ for the setups in Fig. \ref%
{fig1}(a) and \ref{fig2}(a) and $g<3/4$ for the setups of Figs. \ref{fig1}%
(b) and \ref{fig2}(b).  Although in practice it is impossible that the lattice gas collapses to a single site,
 this means the one-dimensional lattice gas with dipole-dipole interaction
 becomes unstable to tend to the center of the lattice. We are unable to estimate the details of the this tendency.
 The time-of-flight experiments measure the momentum distribution of the trapped cloud \cite%
{kett}. In these experiments, the predicted tendency can be observed and
regarded as a ubiquitous evidence of the effect of the renormalized dipole-dipole
interaction.


\subsection{Calculation of the momentum distribution}

\begin{figure}[tbp]
\begin{center}
\includegraphics[width=5.0cm]{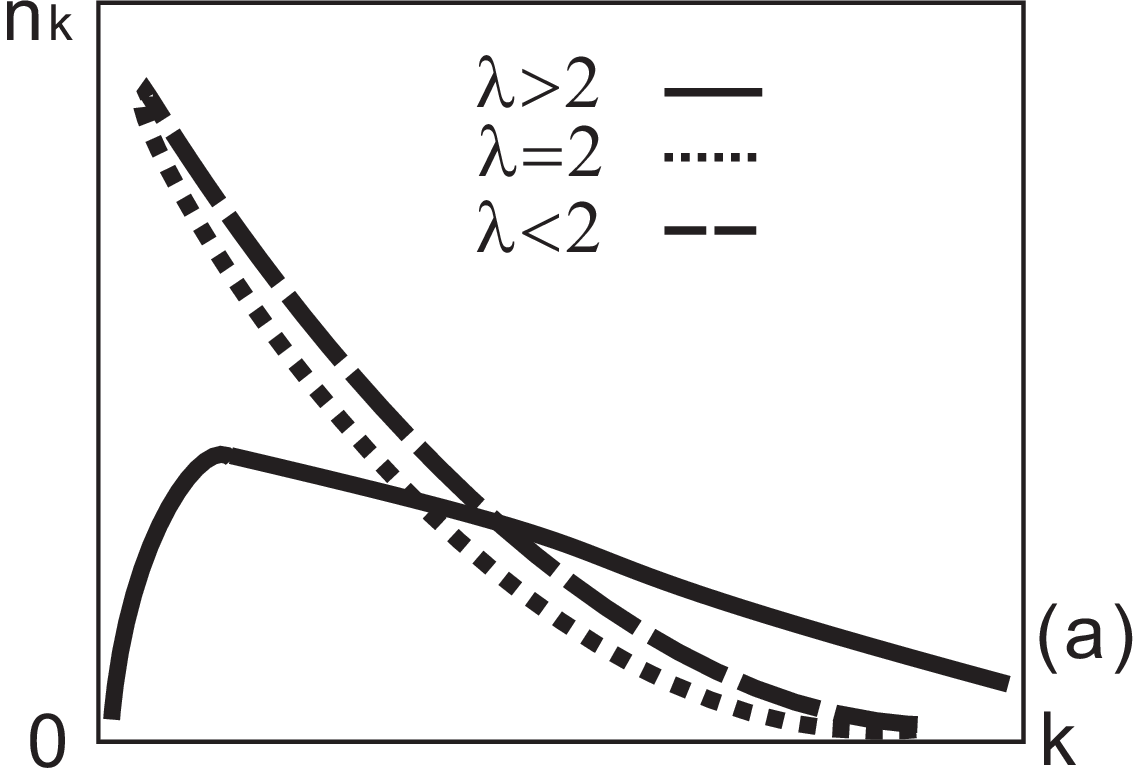}
\par
\includegraphics[width=6.0cm]{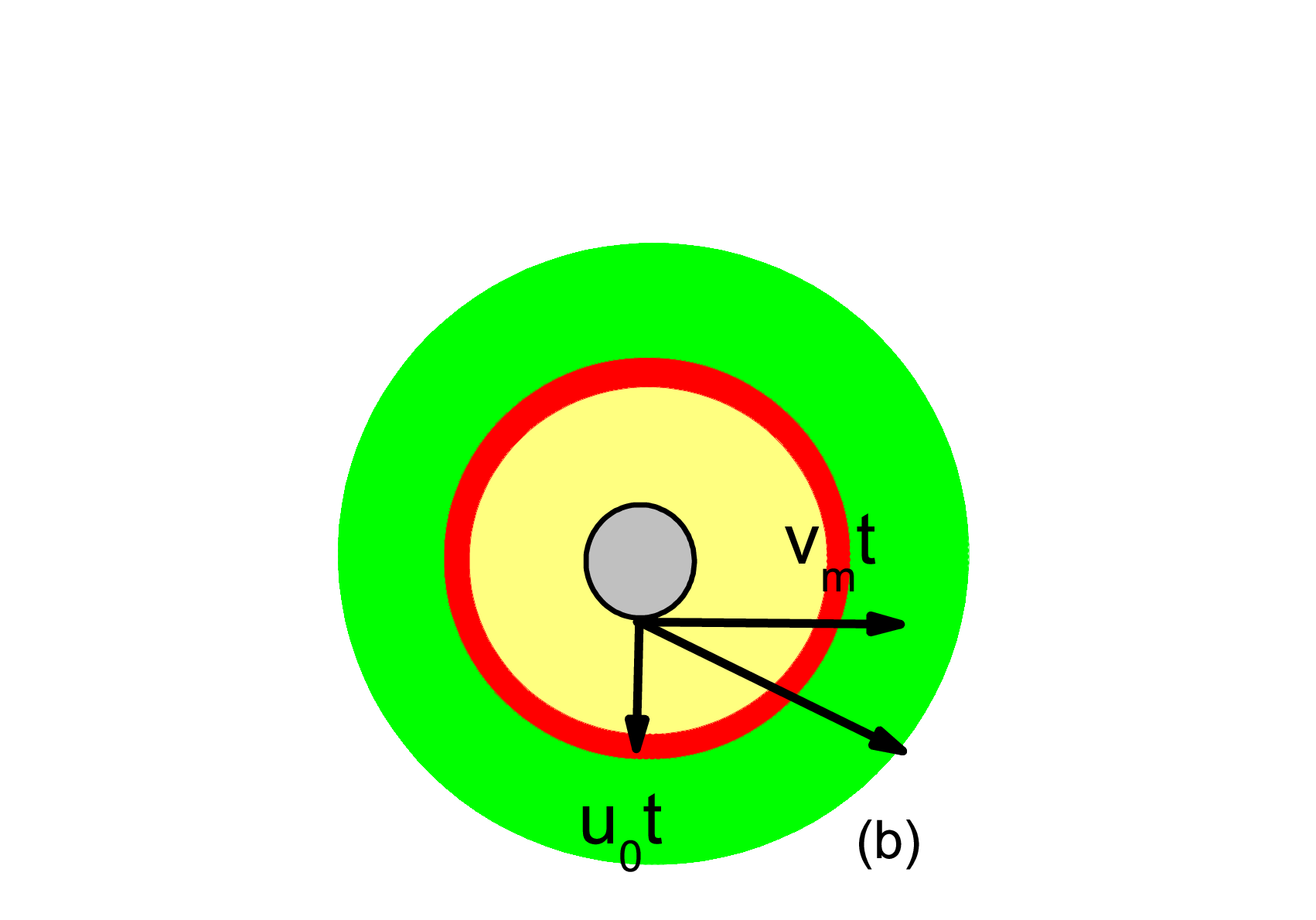}
\end{center}
\par
\vspace{0.2cm}
\par
\begin{center}
\includegraphics[width=5.0cm]{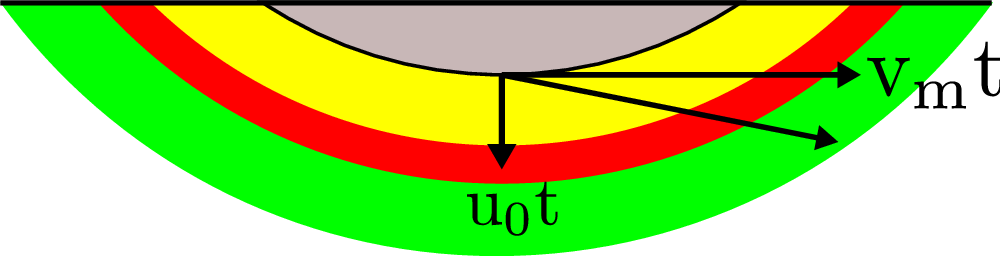}
\end{center}
\par
\vspace{-0.5cm}~~~~~~~~~~~~~~~~~~~~~~~~~~{\small (c)}
\caption{(color online) (a) The schematic diagram of the momentum
distribution. (b) The top-view of the time-of-flight image. The grey area is
the original cylinder. The yellow regime only has very few atoms. Most of
the atoms arrive at time $t$ in the red area and the green region describe
the arrival of those atoms with the momentum distribution in Eq. ( \protect
\ref{gf}). (c) The parabolic counterpart of (b). }
\label{fig3}
\end{figure}

One can study these effects more quantitatively. For example, one can
consider the long wave length limit of the Green's function at a given time
in the Sutherland model, which is given by \cite{Astrakharchik}
\begin{eqnarray}
G(x,0)& \propto&\displaystyle 1+2\sum_{\ell=1}^{\infty} (-1)^\ell B_\ell%
\frac{\cos 2\ell k_F x}{|2k_Fx|^{2\ell^2/\lambda}}  \label{green}
\end{eqnarray}
where $B_\ell$ are regularization-dependent constants and $k_F=\pi\delta
N/L_x$ is the effective Fermi momentum. This corresponds to the infinite
mass or localized limit. The momentum distribution is the Fourier
transformation of this Green's function. For the lowest energy sector with $%
\ell=1$, the momentum distribution in the small $k$ and large $k$ limits are
approximated by \cite{Astrakharchik}
\begin{eqnarray}
n_k&\propto& |k|^{\lambda/2-1}~\mathrm{if~}|k|\ll k_F,  \nonumber \\
&\propto& |k|^{-2\lambda-2}~\mathrm{if~}|k|\gg k_F.  \label{gf}
\end{eqnarray}
We see that when $\lambda<2$, the system is in a quasi-condensed state. When
$\lambda>2$, the momentum distribution vanishes as a power law for small $k$%
, which is the Luttinger liquid behavior for bosons. At the critical point, $%
\lambda=2$, a logarithmic divergence arises, i.e., $n_k=(1/2)\ln (2k_F/|k|)$
for $|k|<2k_F$ and $n_k=0$ for $|k|\geq 2k_F$ \cite{suth}. These momentum
distributions have been shown in Ref. \cite{Astrakharchik} and we display
them schematically in Fig. \ref{fig3}(a).

 Eq. (\ref{gf}) calculated by the Sutherland model
is also valid for the Calogero model because the result is only dependent on
$k_{F}$.

\subsection{Experimental proposal measuring $n_k$}

The projected density profile in the plane perpendicular to the $y$-axis
directly reflects the momentum distribution $n_k$ since the momentum along
the circle of the setup Fig. \ref{fig2} is a good quantum number. The
quasi-condensate is not easy to be distinguished from the case without the
dipolar interaction. However, for $\lambda>2$, when we switch off all traps
but the inner wall of the cylinder, the cloud along the radial direction
expands much faster than that along the axial and tangent directions because
it is the most squeezed along this direction owing to a larger zero point
energy. Denote $u_0$ as the velocity of the radial motion of a single
cigar-shaped cloud and $v_m$ as the fastest velocity of the particles along
the circle, the top view of the cylinder during a time-of-flight experiment
is shown in Fig. \ref{fig3}(b). At time $t$ after turning off the traps,
most of the background BEC atoms arrive at the red area while the
unconventional momentum distribution in Eq. (\ref{gf}) can be measured by
studying the atoms arriving in the green region.

As the setup Fig. \ref{fig2} is not realized in the existent experiments, we
would like to point out that the similar measurement can also be done for
the setup Fig. \ref{fig1}. As we augured, Eq. (\ref{gf}) still hold for the Calogero model.
Because of the existence of the trapping potential, the
time-of-flight process now has a parabolic profile instead of the circular
profile. (See Fig. \ref{fig3}(c). ) When all the traps are removed, we add a
potential above the $z=0$ plane which plays the role of the inner wall
potential in Fig. \ref{fig1} so that the atoms cannot fly upward. The speed $v_m$ of the atoms
flying to the green area is great than $u_0$. Therefore, the density profile in the green area reflects the momentum
distribution in Eq. (\ref{gf}) with $|k|\gg k_F$, i.e., $n_k\propto |k|^{-2\lambda-2}$.

We note that although Eq. (\ref{gf}) provides qualitatively correct results,
more quantitative description of the experimental data can be obtained by
calculating the Green's function (\ref{green}) and the momentum distribution
(\ref{gf}) in terms of the lattice model (\ref{lh}).

\section{Conclusions}

In conclusion, we have shown that the dipole-dipole interaction between the Wannier
modes of cigar-shaped BECs is strongly enhanced and governed by an
inverse-squared potential that dominates over the downward-renormalized
short-range interaction. Based on this finding, we put forth a concrete
proposal to use ultracold alkali atoms on optical lattices as a quantum
simulator to realize the celebrated Calorego-Sutherland model, which in turn provides
insights and predictions to help understand the unconventional many-body
effects in these systems.

\vspace{2mm}

\centerline{\bf ACKNOWLEDGEMENT}

\vspace{2mm}

The authors thank Yingmei Liu, Kun Yang, Su Yi and Li You for useful
discussions. This work was supported by (2012CB821402, 2009CB929101), NNSF
of China (11174298,11121403), and DOE grants DE-FG02-99ER45747 and
DE-SC0002554.


\begin{references}
\bibitem{note} If we take the laser beam wave length $\lambda \sim 850$nm, which gives the lattice spacing $l_p=\lambda/2\sim 425$nm. This yields the tunneling amplitude $t\sim 0.015 E_R$ where $E_R$ is the recoil energy of the atom. This is quite reasonable for a weak interacting optical lattice atom gas. On the other hand, $R_y\sim 10^2\mu$m. The order estimate $l_p/R_y\sim O(10^{-3})$ is reachable.
\end{references}
\end{document}